\documentclass[a4paper,fleqn]{cas-dc}
\usepackage[numbers]{natbib}
\usepackage{lineno,hyperref}
\usepackage{graphicx}
\usepackage{dcolumn}
\usepackage{bm}
\usepackage{siunitx}
\usepackage{braket}
\usepackage{amsmath}
\usepackage{here}
\modulolinenumbers[5]
\begin{document}
\let\WriteBookmarks\relax
\def\floatpagepagefraction{1}
\def\textpagefraction{.001}
\shorttitle{New precise spectroscopy of the hyperfine structure in muonium with a high-intensity pulsed muon beam}
\shortauthors{S. Kanda et~al.}

\title [mode = title]{New precise spectroscopy of the hyperfine structure in muonium with a high-intensity pulsed muon beam}                      

\author[1]{S.~Kanda}[orcid=0000-0002-9080-4154] \cormark[1] \fnmark[1]
\author[2,4,5]{Y.~Fukao}
\author[3,4]{Y.~Ikedo}
\author[1]{K.~Ishida}
\author[1]{M.~Iwasaki}
\author[6]{D.~Kawall}
\author[3,4,5]{N.~Kawamura}
\author[3,4,5]{K.~M.~Kojima} \fnmark[2]
\author[7]{N.~Kurosawa}
\author[8]{Y.~Matsuda}
\author[2,4,5]{T.~Mibe}
\author[3,4,5]{Y.~Miyake}
\author[3,4]{S.~Nishimura}
\author[4,9]{N.~Saito}
\author[2]{Y.~Sato}
\author[1,8]{S.~Seo}
\author[3,4,5]{K.~Shimomura}
\author[3,4,5]{P.~Strasser}
\author[10]{K.~S.~Tanaka}
\author[1,8]{T.~Tanaka}
\author[9]{H.~A.~Torii}
\author[2,4,5]{A.~Toyoda}
\author[1]{Y.~Ueno}
\address[1]{RIKEN, 2-1 Hirosawa, Wako, Saitama 351-0198, Japan}
\address[2]{Institute of Particle and Nuclear Studies, KEK, 1-1 Oho, Tsukuba, Ibaraki, 305-0801, Japan}
\address[3]{Institute of Materials Structure Science, KEK 1-1 Oho, Tsukuba, Ibaraki, 305-0801, Japan}
\address[4]{Japan Proton Accelerator Research Complex (J-PARC), 2-4 Shirakata, Tokai, Ibaraki 319-1195, Japan}
\address[5]{Graduate University of Advanced Studies (SOKENDAI), 1-1 Oho, Tsukuba, Ibaraki 305-0801, Japan}
\address[6]{University of Massachusetts Amherst, 1126 Lederle Graduate Research Tower, Amherst, MA 01003-9337, USA}
\address[7]{Cryogenic Science Center, KEK, 1-1 Oho, Tsukuba, Ibaraki, 305-0801, Japan}
\address[8]{Graduate School of Arts and Sciences, The University of Tokyo, 3-8-1 Komaba, Meguro, Tokyo 153-8902, Japan}
\address[9]{School of Science, The University of Tokyo, 7-3-1 Hongo, Bunkyo, Tokyo 113-0033, Japan}
\address[10]{Tohoku University, 6-3 Aoba, Sendai, Miyagi 980-8578, Japan}

\cortext[cor1]{Corresponding author}
\fntext[fn1]{Present address: Institute of Materials Structure Science, KEK, 1-1 Oho, Tsukuba, Ibaraki 305-0801, Japan.\\
E-mail address: kanda@post.kek.jp}
\fntext[fn2]{Present address: TRIUMF, Vancouver, B. C. V6T 2A3, Canada.}

\begin{abstract}
A hydrogen-like atom consisting of a positive muon and an electron is known as muonium. It is a near-ideal two-body system for a precision test of bound-state theory and fundamental symmetries. The MuSEUM collaboration performed a new precision measurement of the muonium ground-state hyperfine structure at J-PARC using a high-intensity pulsed muon beam and a high-rate capable positron counter. The resonance of hyperfine transition was successfully observed at a near-zero magnetic field, and the muonium hyperfine structure interval of $\nu_{\text{HFS}} = 4.463302(4)\ {\rm GHz}$ was obtained with a relative precision of 0.9 ppm. The result was consistent with the previous ones obtained at Los Alamos National Laboratory and the current theoretical calculation. We present a demonstration of the microwave spectroscopy of muonium for future experiments to achieve the highest precision.
\end{abstract}
\maketitle
\section{Introduction}
Muonium (Mu) is a bound-state of a positive muon and an electron, which was discovered by V.~W.~Hughes $et\ al$~\cite{PhysRevLett.5.63}. In the standard model of particle physics, muonium is a two-body system of structureless leptons. Measurements of muonium's spectral components, such as the 1S-2S interval~\cite{PhysRevLett.84.1136}, the Lamb shift~\cite{PhysRevLett.52.910,PhysRevA.41.93}, and the hyperfine structure (HFS)~\cite{PhysRevLett.82.711} have provided rigorous tests of bound-state quantum electrodynamics (QED) theory and precise determinations of fundamental constants. 

Theoretically, the Mu\,HFS is expressed by the Fermi energy and corrections, including QED, electroweak, and hadronic contributions~\cite{doi:10.1142/0495, KARSHENBOIM20051, IEIDES200163}. The corrections have been collected by the CODATA adjustments of the fundamental physical constants~\cite{RevModPhys.88.035009}. According to a recent re-estimation of the uncertainties, theory predicts~$\nu_{\text{HFS}} \mathalpha{=} 4463.302872(515)$~MHz~\cite{EIDES2019113}. The uncertainty is dominated by the measurement precision of $m_\mu/m_e$ (120~ppb~\cite{PhysRevLett.82.711}), which accounts for 511~Hz out of 515~Hz.

The spectroscopy of the Mu\,HFS yields smaller uncertainty of $\nu_{\text{HFS}}$ than the theoretical prediction. Thus one can obtain a more precise value of $m_\mu/m_e$  by comparing the theoretical prediction and the experimental result. Since this indirectly obtained mass ratio has a much smaller relative uncertainty (20~ppb), it is used to evaluate physical quantities depending on $m_\mu/m_e$. Among them, the muon anomalous magnetic moment $a_\mu$ has been attracting attention because of tension between an experimental result~\cite{PhysRevD.73.072003} and theoretical calculations~\cite{PhysRevD.101.014029,Jegerlehner2018,Davier2020}\footnote[1]{Not all theoretical calculations are inconsistent with the experimental result, $e.g.$, \cite{borsanyi2020leadingorder}. See \cite{aoyama2020anomalous} for a comprehensive review.}.

To measure $a_\mu$ more precisely, a new experiment at Fermi National Accelerator Laboratory (FNAL) is underway \cite{FNAL}, and another one at Japan Proton Accelerator Research Complex (J-PARC) is in preparation~\cite{10.1093/ptep/ptz030}. The uncertainty of $a_\mu$ resulting from the Mu\,HFS is 31 ppb out of 540 ppb  \cite{PhysRevD.73.072003}. This uncertainty is comparable to the major systematic uncertainties expected in the new experiments at FNAL and J-PARC. 

The mass ratio $m_\mu/m_e$ can be obtained from a Mu\,HFS measurement as well as a measurement of the 1S-2S interval in muonium. Recently, new plans for 1S-2S spectroscopy have been proposed \cite{MuMass, Uetake}. Combining results of new measurements of Mu\,HFS and 1S-2S will provide one of the most stringent tests of bound-state QED, and one can extract the Rydberg constant without finite-size effects of a nucleus.

Systems containing second-generation particles amenable to precise spectroscopy are very limited, and thus muonium plays a unique role in searches for physics beyond the standard model and tests of lepton universality. Spectroscopy of the Mu HFS can test the Lorentz invariance with a sidereal oscillation \cite{PhysRevLett.87.111804}, and a search for hypothetical new particles \cite{PhysRevLett.104.220406, PhysRevD.90.073004}. A recently proposed new experiment to measure the Lamb shift in muonium will provide an opportunity for complementary searches \cite{PhysRevD.100.015010, Janka}.

From the 1970s to the 1990s, the Mu\,HFS was measured at the Nevis synchrocyclotron and the Los Alamos Meson Physics Facility (LAMPF). Previous experiments were performed in two ways: observing singlet-triplet transition in a near-zero magnetic field, and measuring the transition frequencies between Zeeman sub-levels in a high magnetic field. The most precise results for each method were $\nu_{\text{HFS}} \mathalpha{=} 4463.3022(14)$~MHz for a zero-field measurement~\cite{CASPERSON1975397}, and $\nu_{\text{HFS}} \mathalpha{=} 4463.302765(53)$~MHz for a high-field measurement~\cite{PhysRevLett.82.711}. The mass ratio was determined with 120~ppb precision.

The previous Nevis experiments were performed using a continuous muon beam incoming at random timing. The scintillation counters detected the muon stopping and subsequent emission of decay positron to measure the time of events. To measure the time difference between muon stopping and positron emission, only one muon per time window of a few microseconds was allowed by the data-acquisition electronics. Therefore, the measurement precision was statistically limited strictly. In the latest experiment at LAMPF, a continuous muon beam was chopped by an electric-field kicker to separate the measurement time window. Approximately $70\%$ of the beam was lost due to this chopping so that statistics limited the measurement precision.

To exceed the limits of previous experiments and realize spectroscopy with higher precision, the MuSEUM collaboration\footnote[3]{MuSEUM is an abbreviation for Muonium Spectroscopy Experiment Using Microwave.} proposed a new experiment using a high-intensity pulsed muon beam at J-PARC~\cite{doi:10.1063/1.3644324}. In contrast to an experiment using a continuous beam, no muon trigger is required because the beam's arrival is synchronized to the accelerator repetition. Bunches of muons are periodically injected, and the Rabi oscillation of muonium is observed as an ensemble average over muonium atoms.

At the Materials and Life Science Experimental Facility (MLF) of J-PARC, the Muon Science Establishment (MUSE) facility delivers the world's highest-intensity pulsed muon beam~\cite{Higemoto_2017}. However, its benefit involves difficulties in positron counting due to the high instantaneous event rate. The novelty and significance of the experiment described in this paper are the high-intensity pulsed muon beam's application to precise spectroscopy using a high-rate capable particle detector.

\section{Theory}
The Mu HFS $A \mathalpha{=} h \nu_{\text{HFS}}$ of muonium in the ground-state is the energy splitting between the spin-triplet state and the spin-singlet state. The Hamiltonian of muonium in a magnetic field is described as
\begin{equation}
\mathcal{H}=A \bm{S}_\mu \cdot \bm{S}_e + (g'_e \mu^{\text{B}}_e \bm{S}_e - g'_\mu \mu^{\text{B}}_\mu \bm{S}_\mu ) \cdot \bm{B},
\end{equation}
where $\bm{S}_l$ is the spin operator of muon or electron ($l \mathalpha{=} \mu, e$, the same shall apply hereinafter), $g'_l$ is the bound-state $g$-factor in muonium~\cite{PhysRevA.4.59} \footnote[4]{Here the sign of the $g$-factor is defined as positive.}, $\mu^{\text{B}}_l \mathalpha{=} e \hbar / 2 m_l$, $m_l$ is the mass, and $\bm{B}$ is the external magnetic field.

Microwave irradiation at an appropriate frequency excites muonium from the singlet-state to the triplet-state. The associated time-dependent Hamiltonian is represented as
\begin{equation}
\mathcal{H}_\text{I}(t) =  ( g'_e \mu^{\text{B}}_e \bm{S}_e - g'_\mu \mu^{\text{B}}_\mu \bm{S}_\mu ) \cdot \bm{B}_{1} \cos \omega t.
\end{equation}
Here $\bm{B}_{1}$ is the magnetic field of the applied microwave, and $\omega$ is its angular frequency.

When the external magnetic field is sufficiently weak, the energy eigenstates of muonium are classified by the total angular momentum $F$ and the associated magnetic quantum number $m_F$ as $(\psi_1,\psi_2,\psi_3,\psi_4)=(\ket{1,1},\ket{1,0},\ket{1,-1},\ket{0,0})$, where the first and the second number indicates $F$ and $m_F$, respectively. The Hamiltonian based on the energy eigenfunctions of muonium is explicitly written as
{\setlength\arraycolsep{2pt}
\begin{eqnarray}
\hspace{-1cm}
\mathcal{H}' &=& \mathcal{H} + \mathcal{H}_\text{I}(t) \nonumber \\
&=&\hbar
\begin{pmatrix}
\Omega_\text{L} & 0 & 0 & 2\Omega_\text{R} \cos \omega t \\
0 & 0 & 0 & 0\\
0 & 0 & - \Omega_\text{L} & - 2\Omega_\text{R} \cos \omega t \\
2\Omega_\text{R} \cos \omega t & 0 &  -2\Omega_\text{R} \cos \omega t & -2 \pi  \nu_{\text{HFS}}
\end{pmatrix},
\end{eqnarray}}
where $\Omega_\text{L}  \mathalpha{=} (g'_e \mu^{\text{B}}_e  \mathalpha{-} g'_\mu \mu^{\text{B}}_\mu) {B}/{2\hbar}$ is the Larmor frequency, $B$ is the static field strength along the $z$-axis, $\Omega_\text{R}  \mathalpha{=} ( g'_e \mu^{\text{B}}_e  \mathalpha{+} g'_\mu \mu^{\text{B}}_\mu ) {B_1}/{(4\sqrt{2} \hbar)}$ is the Rabi frequency, and $B_1$ is the microwave field strength.

The Rabi oscillation between the hyperfine-states causes a time-evolution of the muon spin polarization, correlating with decay positrons' emission angle. Therefore, an experimental observable is oscillating positron counts associated with the Rabi oscillation.
The signal in the experiment $S(t)$ is defined by taking the ratio of positron counts with ($N_{\text{ON}}(t)$) and without ($N_{\text{OFF}}(t)$) microwave irradiation,
\vspace{-0.2cm}
\begin{equation}
\label{eq:rabi}
S(t) = \frac{N_{\text{ON}}(t)}{N_{\text{OFF}}(t)} - 1.
\end{equation}
The signal defined by above will be denoted the Rabi-oscillation signal hereafter. The ratio of positron counts integrated over time yields the resonance curve as a function of the varying microwave frequency.

A theoretical expression of the signal is derived from calculating the state amplitudes using the density matrix for a statistical mixture of muonium states. The theoretical expressions of the Rabi oscillation and the resonance curve are obtained in the references~\cite{PhysRevA.3.871, PhysRevA.5.2338, PhysRevA.8.86}. The signal at a certain microwave field strength $L(t)$ is written as follows,
\begin{equation} 
\hspace{-0.2cm}
\label{eq:Lt}
L(t) =  \Big[  \frac{\Gamma \mathalpha{+} \Delta \omega}{\Gamma} \cos\frac{\Gamma \mathalpha{-} \Delta \omega}{2}t \mathalpha{+} \frac{\Gamma \mathalpha{-} \Delta \omega}{\Gamma} \cos\frac{\Gamma \mathalpha{+} \Delta \omega}{2}t \Big] s \text{e}^{ - \lambda t},
\end{equation}
where $\Gamma  \mathalpha{=} \sqrt{ \Delta \omega^2  \mathalpha{+} 8 \Omega_\text{R}^2}$, $\Delta \omega  \mathalpha{=} \omega  \mathalpha{-} 4463.303  \mathalpha{\times} 2\pi$~MHz is the microwave frequency detuning, $s$ is the scaling factor depending on the acceptance of the positron detector and the minimum energy of detected positrons, and $\lambda$ is the damping constant, which represents muon spin depolarization. Depolarization may occur due to magnetic impurities in gas, such as oxygen. Since the microwave field's strength is position-dependent, the signal is observed as the sum of multiple oscillation components.

\section{Experiment}
The experiment was conducted at J-PARC MLF MUSE D-Line. A pulsed 3~GeV proton beam was injected into a graphite target, and hadronic interactions produce pions. The decay of pion at rest on the target surface yielded spin-polarized positive muon ($\mu^+$). A muon beam having a momentum of 27.4~MeV/$c$ irradiated krypton gas at a pressure of 1~bar to form muonium atoms after muon stopping in the gas target. The beam intensity was $2\times10^6~\mu^+$/s with the accelerator operation power of 200~kW. The beam was pulsed and repetitive at 25 Hz, resulting in $8\times10^4~\mu^+$ per pulse. The momentum spread of the beam was $\Delta p/p=10\%$ (FWHM).

\begin{figure}[h]
\centering
\includegraphics[width=\linewidth]{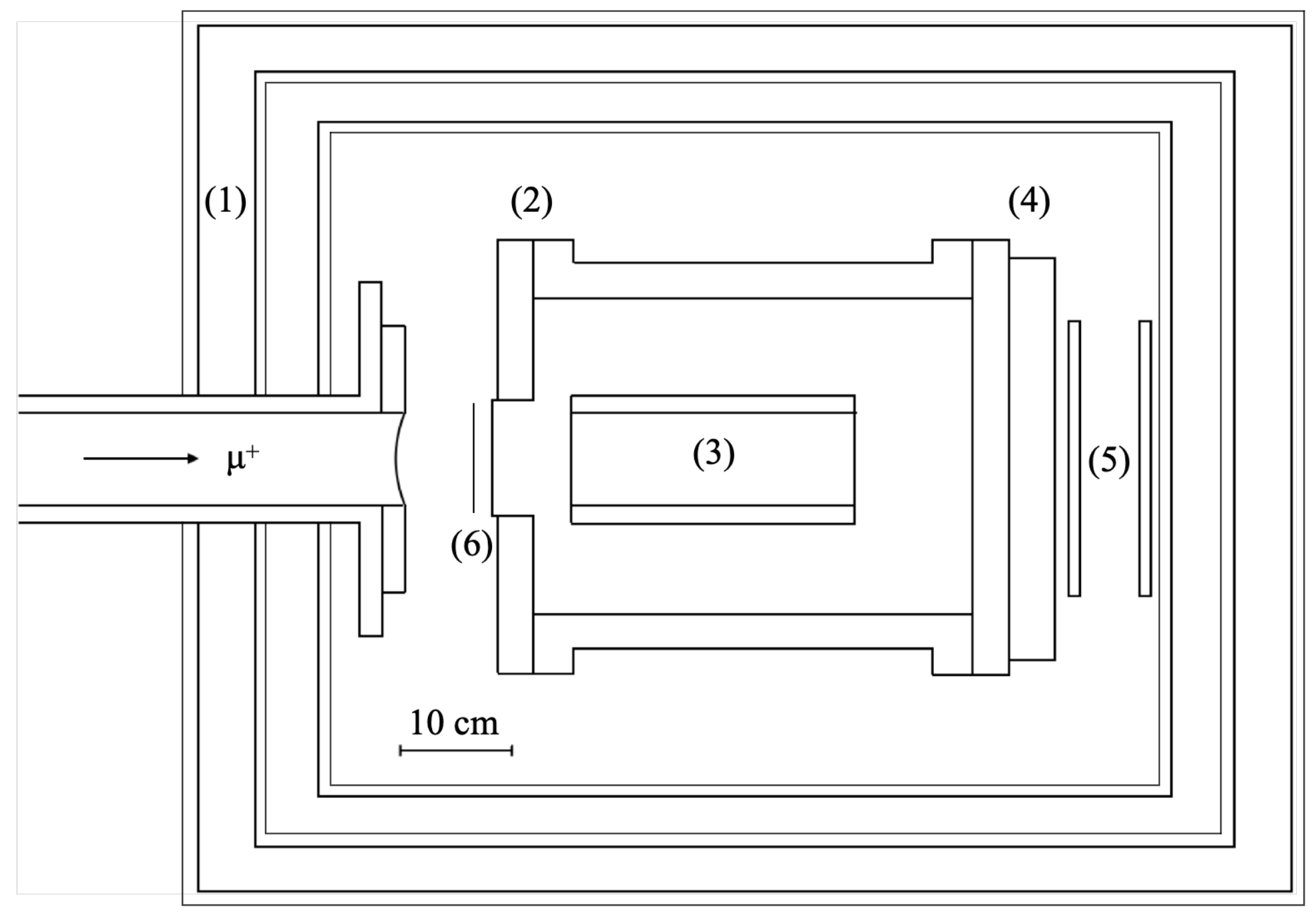}
\vspace{-0.2cm}
\caption{Drawing of the experimental apparatus: (1) three-layers of the magnetic shield, (2) the cylindrical gas chamber made of aluminum, (3) the cylindrical microwave cavity made of copper, (4) the aluminum absorber for background suppression, (5) the segmented positron counter, (6) the fiber hodocsope.}
\vspace{-0.5cm}
\label{fig:exp}
\end{figure}

Figure \ref{fig:exp} illustrates the experimental setup. Krypton gas with the purity of $99.999\%$ was confined in a cylindrical aluminum vessel with an inner diameter of 280~mm and an axial length of 450~mm. The gas pressure was measured by a capacitance gauge (ANELVA M-342DG) with $0.2\%$ accuracy. The chamber's upstream end had a thin aluminum beam window with a thickness of 100~${\rm \mu m}$ and a diameter of 100~mm. At the beam window, the muon beam profile was measured by a fiber hodoscope. The profile was a two-dimensional Gaussian with a standard deviation of 2~cm.

A Monte-Carlo simulation using GEANT4 toolkit~\cite{GEANT4_1,GEANT4_2,GEANT4_3} was performed, and the fraction of muon stopping in the cavity was estimated to be $30\%$ of the total incident. Almost all muons stopped in krypton gas become muonium.

The initial state population of muonium is statistically distributed in the spin-singlet state (25$\%$) and the spin-triplet states (75$\%$)~\cite{PhysRevA.1.595}. Irradiation of microwave induces transitions between the states. This hyperfine-state transition causes muon spin flip. The time evolution of the muon spin was observed via the angular asymmetry of positrons from muon decays. A segmented plastic scintillation counter detected the decay positrons.

\begin{figure}[h]
\vspace{-0.1cm}
\centering
\includegraphics[width=0.6\linewidth]{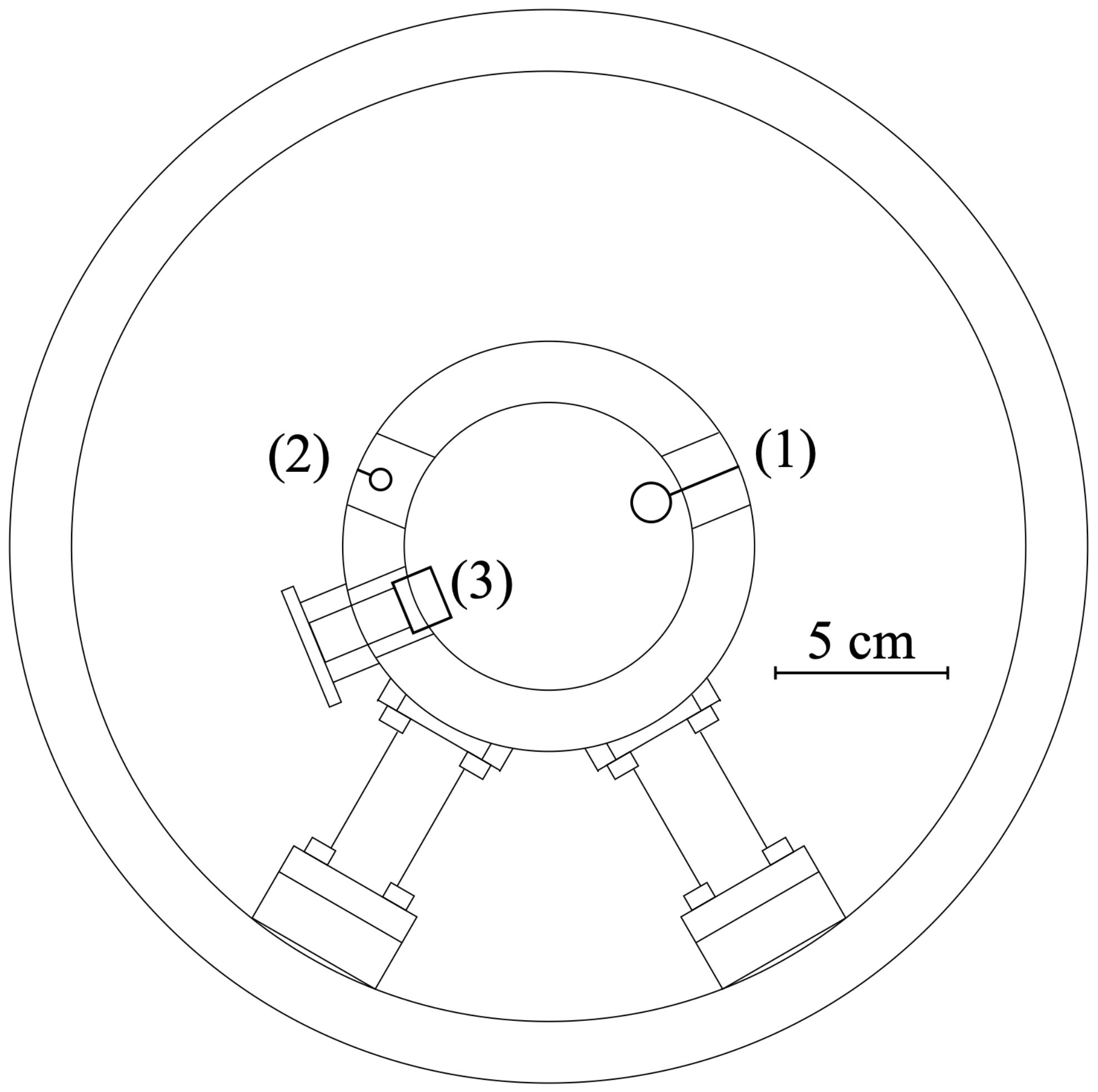}
\caption{Drawing of the microwave cavity: (1) loop antenna for input, (2) loop antenna for power monitoring, (3) aluminum tuning rod with the piezoelectric actuator.}
\label{fig:cavity}
\end{figure}

A cylindrical cavity made of oxygen-free copper with an inner diameter of 81.8~mm was used to apply microwaves to the muonium atoms. Figure \ref{fig:cavity} shows the drawing of the cavity. An inner axial length of the cavity was 230~mm so that muons could be sufficiently stopped in the gas target. The microwave resonated in TM110 mode with a quality factor of 5000 at 4463.302~MHz. The quality factor was frequency-dependent, as Fig. \ref{fig:q} presents in the appendix. The microwave from a signal generator (Hewlett-Packard 8671B) was input to the cavity through amplifiers (Mini Circuit ZVE-8G). The microwave power was monitored by a thermal power sensor (Rohde\&Schwarz NRP-Z51), and typical input power was about 0.85 W. The resonance frequency was tuned by moving an aluminum rod inserted into the cavity with a piezoelectric actuator (attocube ANPz101eXT12). The frequency ranged from 4461.8~MHz to 4464.8~MHz. The microwave was switched at ten-minute intervals to suppress the temperature rise of the cavity.
\vspace{-0.3cm}

A three-layer box-shaped magnetic shield made of an alloy of iron and nickel was used against the geomagnetic field and the static magnetic field generated by surrounding devices. The three-dimensional magnetic field distribution in the cavity was measured by a fluxgate probe (MTI FM3500) with 0.5 nT resolution. The static magnetic field inside the cavity was less than 60~nT. A compact air conditioner (ORION PAP01B) was employed to keep the temperature inside the shield constant.

\begin{figure}[h]
\centering
\includegraphics[width=\linewidth]{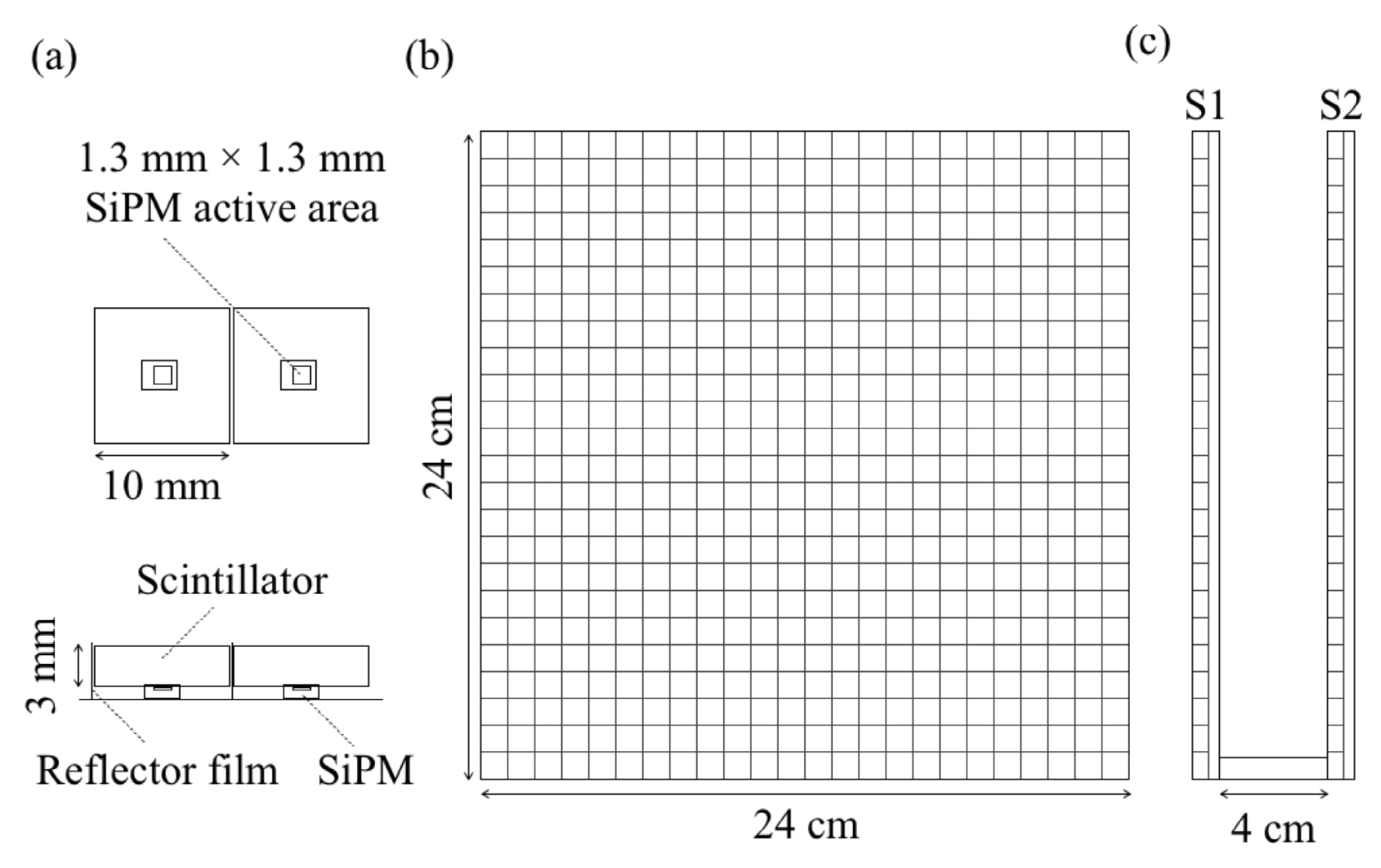}
\caption{Drawing of the positron counter: (a) enlarged view of the scintillator tile and SiPM, (b) overall view seen from the beam, (c) view from the side.}
\label{fig:det}
\end{figure}

The segmented positron counter consisted of an array of plastic scintillator tiles and silicon photomultipliers (SiPMs) \cite{Kanda:2016fuz}. Figure \ref{fig:det} depicts the positron counter. The detector had two layers, 4~cm apart, consisting of 24-by-24 scintillator tiles. A SiPM (Hamamatsu Photonics MPPC S12825-050P-01) with an active area of 1.3~mm square was connected to each scintillator (Eljen Technology EJ-212). The scintillator tiles of 1~cm square and 3~mm thick were two-dimensionally arranged. Reflector films (3M ESR) were inserted between tiles. The upstream layer of the detector was placed 20~cm away from the downstream end of the cavity. The geometrical acceptance considering multiple scattering through the materials was estimated to be $1\%$.

The signals from the SiPMs were processed by the Kalliope front-end electronics consisting of an ASIC\footnote[5]{Application Specific Integrated Circuit}-based amplifier-shaper-discriminator and a multi-hit time-to-digital converter implemented in FPGA\footnote[6]{Field Programmable Gate Array}, where the leading edge time was recorded \cite{Kojima_2014}. The photon yield of a positron from muon decay was represented by a Landau distribution with a peak at 55 photons.
The discriminator threshold was set at 1.5 photon equivalent (p.e.)~level so that the detection efficiency for incident positrons was almost 100$\%$. The typical dark count rate of each SiPM at 1.5 p.e.\ threshold was 17~kHz. The time resolution of the detector was 8~ns~(1$\sigma$). 
 
A large number of prompt positrons from the muon production target and positrons from muon decay during transport were incident on the apparatus. The momentum of these background positrons was similar to that of the transported muons, 27.4~MeV/$c$. To prevent these positrons from causing background events, an aluminum plate with a 40 mm thickness was placed between the target chamber and the detector. This plate served as an absorber to block positrons with momentum below 40 MeV/$c$. The background events were suppressed by a factor of five. Besides, the energy threshold selected the positrons emitted preferentially along the muon spin direction. The loss due to the absorber was estimated by a GEANT4 simulation to be $40\%$. 

\section{Analysis}
\begin{figure}[h]
\centering
\includegraphics[width=\linewidth]{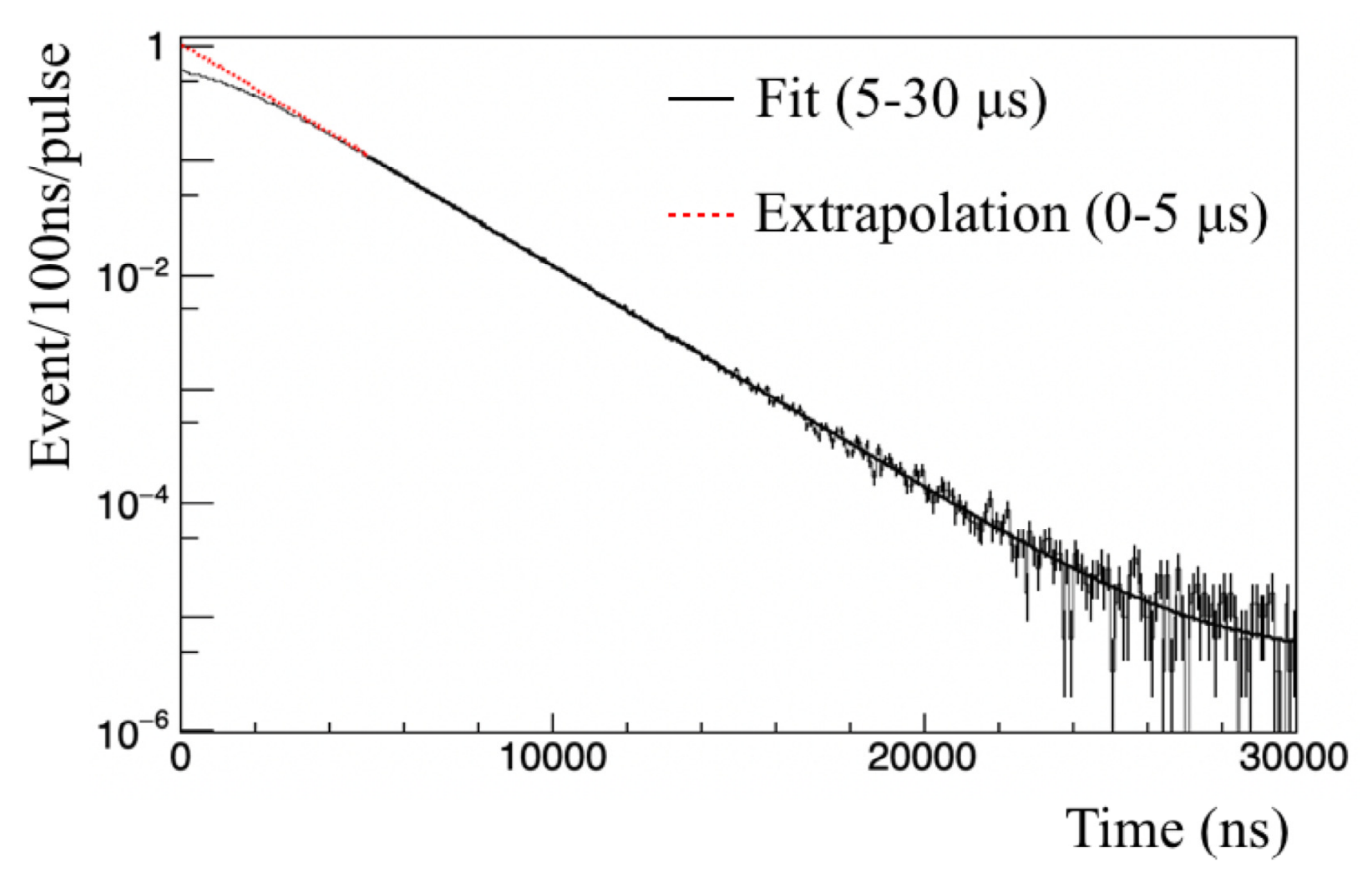}
\caption{Time spectrum of the number of decay positrons and background events without microwave irradiation. The number of beam pulses normalizes the ordinate. The black solid curve shows the fitting result with an exponential function on a constant background. The fitting exponent gives the muon lifetime of 2198(5) ns. The red dashed line indicates the extrapolation of the fitting function.}
\label{fig:spectrum}
\end{figure}

Data for 15 hours of measurement was analyzed. In data analysis, the background events due to dark counts of SiPMs were suppressed by selecting coincidence events found in the two detector layers. The time window of the coincidence analysis was set to 24 ns, which corresponded to three times the time resolution. Simultaneous hits in adjacent segments on the layer were merged as a hit-cluster having the same origin. Figure~\ref{fig:spectrum} shows the time spectrum. In an ideal situation without pileup counting loss, the spectrum is exponential with the muon mean lifetime. 

\begin{figure}[h]
\centering
\includegraphics[width=\linewidth]{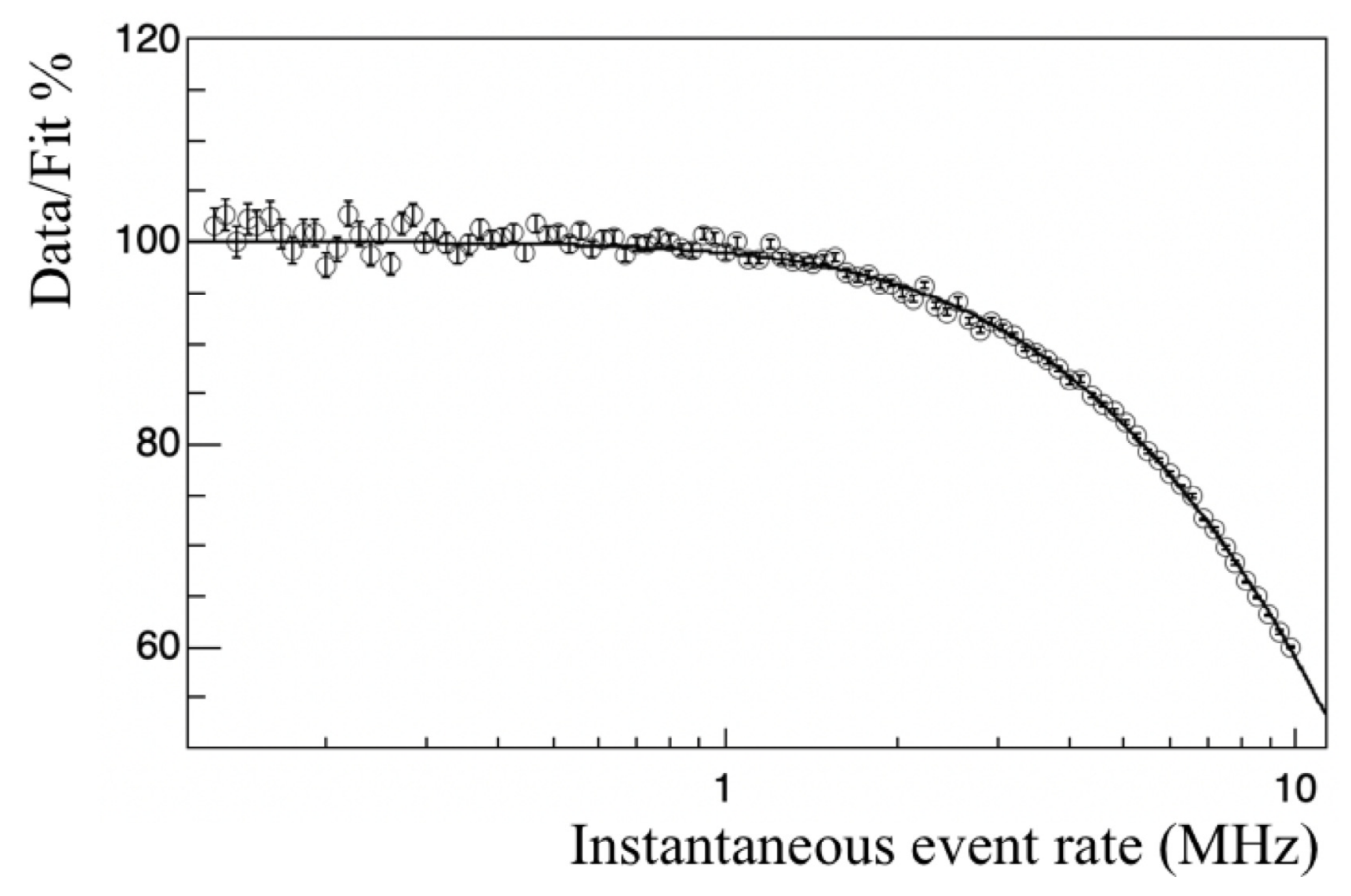}
\caption{Pileup counting loss as a function of the instantaneous event rate. The loss was calculated from the ratio of the data points and the function in Figure \ref{fig:spectrum}. The black curve indicates the result of fitting with the model function \cite{Ida:hx5024}.}
\label{fig:pileup}
\end{figure}

A bunch of pulsed muons makes multiple hits on the detector simultaneously. The simultaneous overlap of multiple positrons causes signal counting loss. This pileup event occurs more frequently when the instantaneous count rate is higher. The pileup effect was evaluated by taking the difference between the observed spectrum and the extrapolated fitting result obtained in the low rate region, where pileup loss is negligible. Figure~\ref{fig:pileup} shows the relative efficiency considering pileup loss.
The measurement result is well explained by a pulse-height analyzer (PHA) windowing model \cite{Ida:hx5024}. The detector's dead-time obtained by the fitting analysis was 500 ns, consistent with the analog signal observations. The counting loss due to pileup was about $20\%$ of the total detection. The number of coincidence events per beam pulse was about 110, consistent with the expectation considering pileup loss. The detector's rate capability needs to be high enough so that the pileup counting loss should not cause serious systematic uncertainty. Details will be given in the next section.

\begin{figure}[h]
\centering
\includegraphics[width=\linewidth]{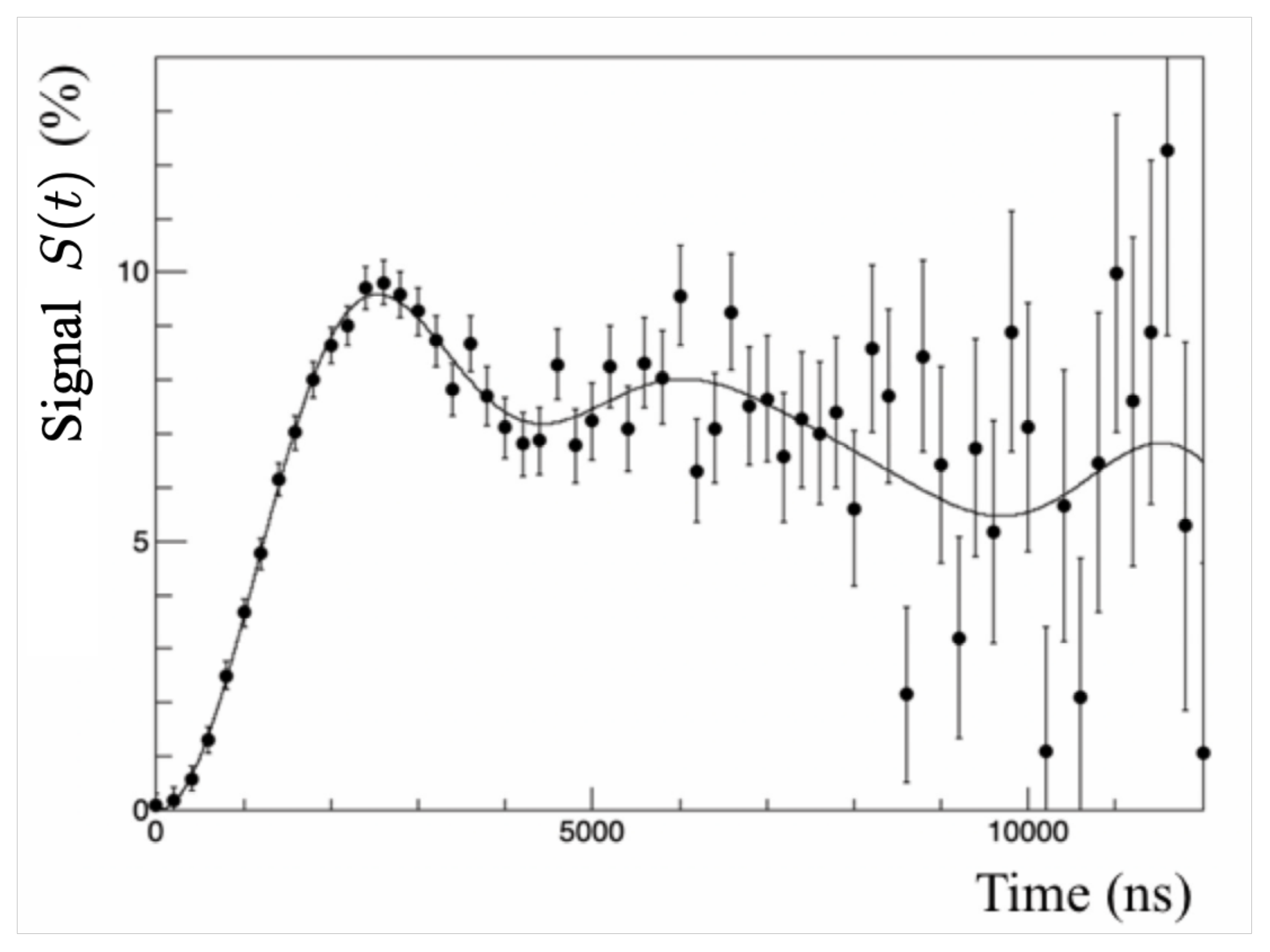}
\vspace{-0.1cm}
\caption{The Rabi oscillation of muonium under the microwave field with a frequency of 4463.302 MHz. The solid curve shows the fitting result using the theoretical expression of the signal defined by Eq. (\ref{eq:Lt}). The frequency detuning was fixed to zero in the fitting. Eight oscillation components corresponding to representative microwave strengths were assumed. The reduced chi-square is $\chi^2/{\rm NDF} = 50/45$, which gives the $p$-value of 0.28.}
\label{fig:rabi}
\vspace{-0.5cm}
\end{figure}

The Rabi-oscillation signal was obtained by taking the ratio between time spectra with and without microwave irradiation as defined by Eq. (\ref{eq:rabi}). Figure \ref{fig:rabi} shows the result with the microwave frequency of 4463.302~MHz. While the fitting function included the spin depolarization time-constant as a parameter, no significant depolarization was observed.

\section{Result}
\begin{figure}[h]
\centering
\includegraphics[width=\linewidth]{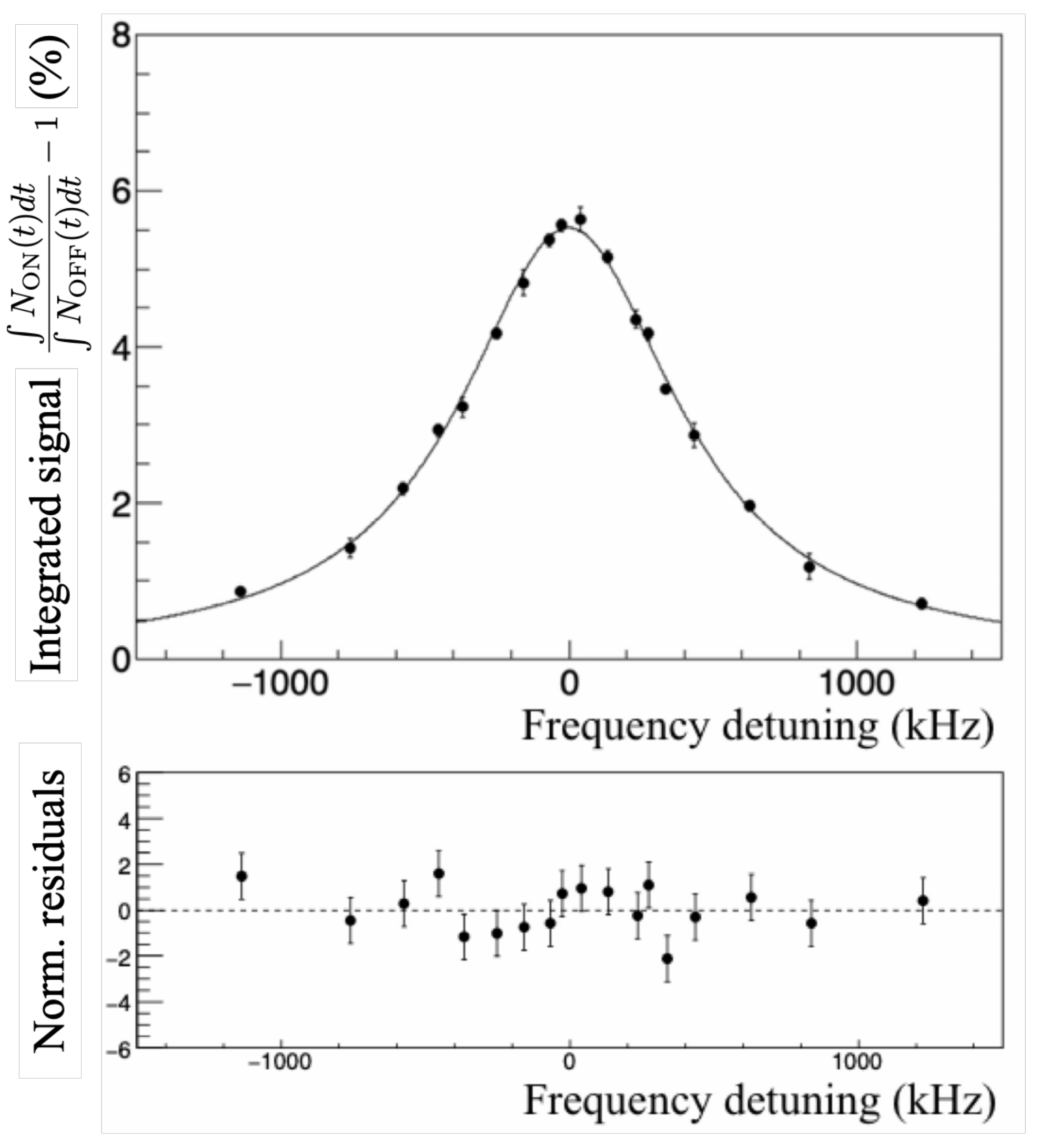}
\caption{Result of the frequency scan measurement. The upper panel shows the resonance curve. The vertical axis corresponds to the time integration of the Rabi-oscillation signal. The horizontal axis represents the frequency detuning from 4463.302~MHz. The solid curve shows the fitting result with a Lorentz function. The reduced chi-square is $\chi^2/{\rm NDF} = 16.9/15$, which gives the $p$-value of 0.32. The normalized fitting residuals are shown in the lower panel.}
\label{fig:result}
\end{figure}

The ratio of integrals of positron spectra with and without microwave yields a resonance curve by sweeping the microwave frequency. Figure \ref{fig:result} shows the resonance curve as a result of the experiment. The frequency dependence of the cavity quality factor was corrected, as described in the appendix. Density-dependent frequency shift due to atomic collisions \cite{PhysRevA.2.1411} was corrected for using the past experimental result, which amounted to the shift of 36 kHz at the krypton gas pressure of 1.013(2) bar \cite{PhysRevA.8.86}. As will be described later, the systematic uncertainty associated with this correction was 46 Hz. The resonance frequency was determined by fitting the curve with a Lorentz function. In this analysis, the microwave field strength was represented by one parameter instead of field distribution, similar to the previous study in a near-zero field \cite{CASPERSON1975397}. The analysis gave the Mu HFS of
\begin{equation}
\nu_{\text{HFS}} = 4463.302(4)~\text{MHz},
\end{equation}
where the uncertainty is statistical. The result was consistent with the theoretical prediction and the previous experimental results with the continuous muon beams. No significant systematic deviation from the fitted curve was observed.

The systematic uncertainties in the experiment are summarized in Table \ref{table:syst}. The total uncertainty was calculated, assuming each contribution was independent. At present, no significant correlation of systematic uncertainties is found. With the present apparatus, the systematic uncertainty was dominated by the pressure gauge's absolute accuracy, making the density shift correction ambiguous (46~Hz). The second-largest contribution was due to the instability of the microwave power (37~Hz). Both systematic uncertainties were far less than the statistical uncertainty (4~kHz). In a future experiment, quantitative evaluation of the systematics will be essential after completing a brand-new beamline construction that enables long-term measurement.

Using the Rabi-oscillation signal and the pileup model function, the systematic effect of pileup loss on the resonance curve was evaluated. While the analysis using the spectral ratio mostly canceled the impact of the pileup, a small time-dependent change in the counting rate accompanying the oscillation affected the resonance curve. The systematic uncertainty was numerically evaluated to be 19~Hz, sufficiently small compared to the statistical uncertainty.

Fluctuations in gas temperature affect the correction of density dependence. The two major causes of fluctuation were heat dissipation from the cavity and changes in the ambient temperature. These two effects were clearly observed from the gas pressure measurement. During the experiment, the gas temperature change was estimated to be 0.1~K for the rise of the cavity temperature and 0.2~K for the ambient temperature change.

The contribution from the static magnetic field was evaluated by calculating the effect of Zeeman splitting on the resonance curve. Impurities in the gas target were quantified with a quadrupole mass spectrometer, and the effect of spin relaxation was evaluated from the upper limit of oxygen partial pressure (0.4 ppm, after 20 hours of gas exchange). The stability of the muon beam intensity was evaluated using proton beam intensity obtained by a current transformer. The fluctuation of the muon beam profile affects the effective strength of the microwave field. The expected variation of the profile was evaluated by the current stability of the bending magnet in the beamline being $0.01\%$.
\begin{table}[hbtp]
\caption{Systematic uncertainties in the experiment.}
\label{table:syst}
\centering
\begin{tabular}{l c}
\hline
Source & Contribution (Hz)\\
\hline
Gas density measurement& 46\\
Microwave power drift& 37\\
Detector pileup & 19\\
Gas temperature fluctuation& 6\\
Static magnetic field & $\text{negligible}$\\
Gas impurity buildup & 12\\
Muon beam intensity & $\text{negligible}$\\
Muon beam profile & $\text{negligible}$\\
\hline
Total & 63\\
\hline
\end{tabular}
\end{table}
 
In the experiment, a statistical uncertainty of 4~kHz was obtained for 15~hours of measurement. The expected beam intensity of the brand-new beamline under construction is $1\times10^8~\mu^+$/s \cite{10.1093/ptep/pty116}. This corresponds to an intensity improvement of 50 times. In a zero-magnetic field experiment using the present apparatus and the new beamline, the statistical precision will be comparable to the previous result in 72 days of measurement.

In a high-field experiment, the muon spin polarization after muonium formation becomes 100$\%$, and more decay positrons reach the detector by focusing along the longitudinal field. In addition, the diameter of the cavity becomes larger for a lower resonance frequency due to the Zeeman shift. These effects are evaluated by simulations using GEANT4. The statistical precision will reach 5~Hz (1.2~ppb) in 40~days of measurement. This precision corresponds to ten times better than the previous experiment. A similar improvement is expected for the muon-to-electron mass ratio; that is, the mass ratio can be determined with 12~ppb precision. This is comparable to the goals of the future measurements of the 1S-2S interval in muonium \cite{MuMass, Uetake} \footnote[8]{For the Mu-MASS experiment, this precision corresponds to the Phase-1 goal.}.

The higher the beam intensity, the higher the rate of pileup. However, the count rate can be reduced without compromising the statistical power by the "old muonium" method, which selectively analyzes muonium atoms that lived longer than the muon lifetime. This technique can be used for both zero- and high-field experiments \cite{CASPERSON1975397,PhysRevA.52.1948}. Even in a high-field, which gives a higher rate than a zero-field case, the count rate becomes acceptable with a delay of three times the lifetime.

For future high-precision experiments, a reference pressure monitor (FLUKE RPM4 A1.4Ms) is prepared. The accuracy of measurement is 0.01$\%$ at 0.35 bar, which is 20 times better than the capacitance gauge. For microwave power stabilization, pulse-by-pulse switching of microwave and water cooling of the cavity are in preparation. With these measures, the uncertainty arising from the microwave power drift is expected to be negligible. Water cooling of the cavity will also suppress the temperature fluctuation. A more precise evaluation of magnetic impurities is in preparation. 

\section{Conclusion}
New precision measurement of the Mu\,HFS using the high-intensity pulsed muon beam was performed at J-PARC MLF MUSE. This measurement is a milestone for a future long-term measurement with a further enhanced beam-intensity to surpass the past results. The measurement principle was proven under a precisely-controlled near-zero magnetic field. The segmented positron counter was employed to maximize the advantage of the high-intensity beam. The result was $\nu_{\text{HFS}} = 4463.302(4)~\text{MHz}$ (0.9 ppm).
A zero-field experiment using the brand-new beamline will achieve a comparable precision (12 ppb) as the most precise result in the past after 72 days of measurement. In a high-field experi- ment aiming for the highest precision, ten-fold improvement (1.2 ppb) is expected in 40 days, considering the beam intensity and effects of a high magnetic field on the measurement.

\section*{Acknowledgements}
The authors thank the J-PARC and KEK personnel for scientific, engineering, and technical supports to prepare and accomplish the experiment. This experiment was conducted under the user program  2017A0134 at MLF of J-PARC. This work was supported by Japanese JSPS KAKENHI Grant Numbers 23244046, 26247046, 14J11374, 15K17666, 15H05742 and 17H01133.

\appendix
\section{Correction of the cavity quality factor}
In the experiment, the power input to the cavity was set constant. The strength of the microwave field $|B_1|$ is proportional to $\sqrt{Q/\omega}$, where $Q$ is the cavity quality factor, and $\omega$ is the frequency \cite{PhysRevA.8.86}. The frequency dependence is negligibly small rather than quality-factor dependence. The Rabi frequency is proportional to $\sqrt{Q}$. In the analysis, the time-integrated signals were calculated with and without the frequency dependence of the $Q$. The correction factor was obtained from the signal ratio at each frequency, and it was almost linear with $Q$.
\begin{figure}[h]
\centering
\includegraphics[width=0.8\linewidth]{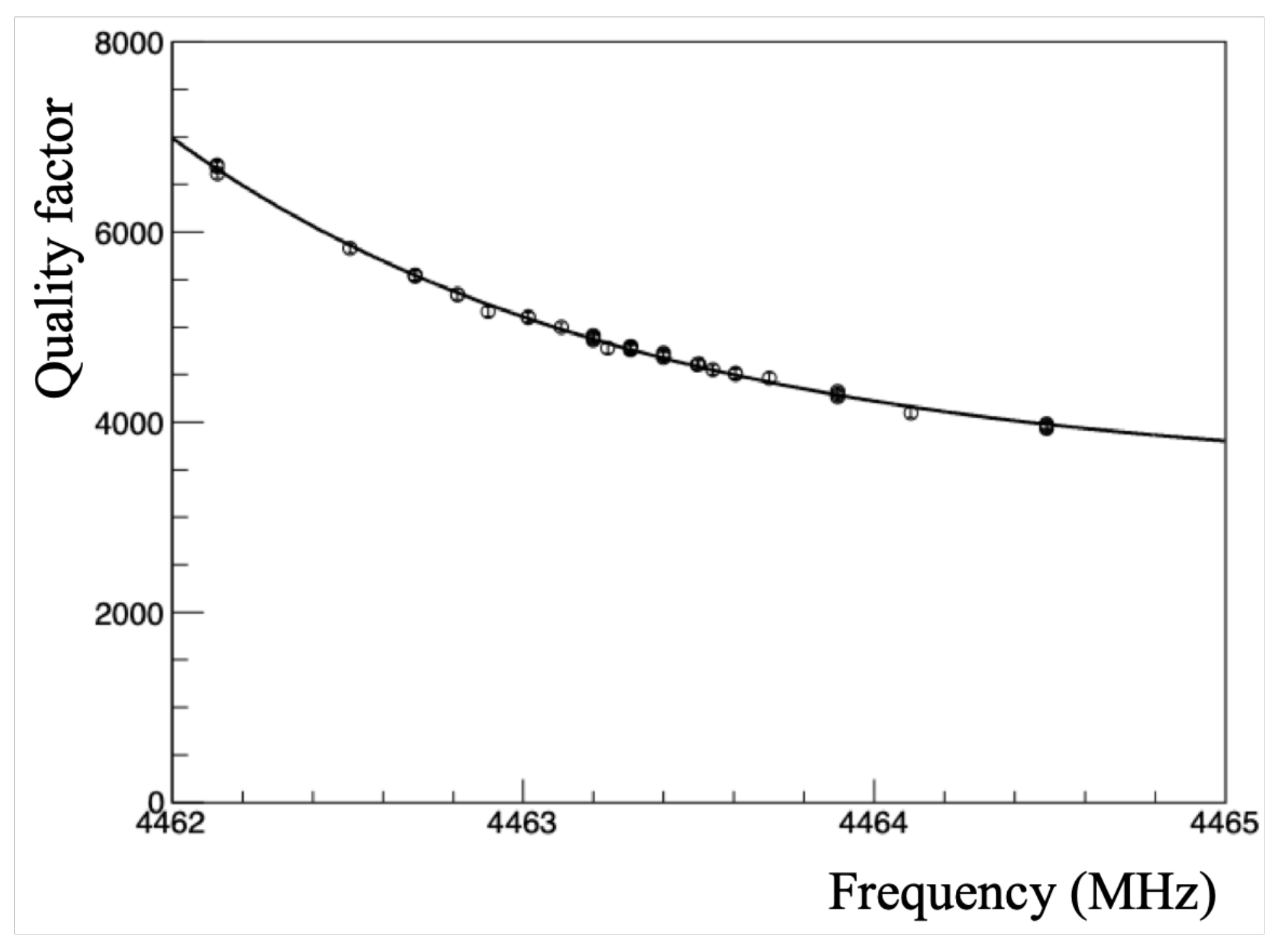}
\caption{The cavity quality factor as a function of the resonance frequency. The solid curve represents the fitting result with an exponential function.}
\label{fig:q}
\end{figure}

\bibliographystyle{model1a-num-names.bst}
\bibliography{mybibfile}
\end{document}